%% file: latex_template_MedPhys_2021.tex
\documentclass[12pt,twoside]{article}   
\usepackage[super,sort&compress,comma]{natbib}

\usepackage{subcaption}
\usepackage{fancyhdr}		

\newcommand{\captionv}[3]{\begin{center}\parbox{#1cm}{\caption[#2]{{\sf #3}}}
        \end{center}}

\usepackage[section]{placeins}   %

\usepackage{graphicx}
\usepackage{siunitx}
\usepackage{xspace}
\usepackage{makecell}
\usepackage{caption}
\usepackage{float}
\usepackage{setspace}
\usepackage{xspace}
\usepackage{soul}
\makeatletter \renewcommand\@biblabel[1]{$^{#1}$} \makeatother
\renewcommand{\thesubsection}{\thesection{\sf \Alph{subsection}}.}

\usepackage[labelfont=bf, labelsep=period]{caption}

\newcommand{\ie}{i.e.,\xspace}

\newcommand{\cen}[1]{\begin{center} #1 \end{center}}

\usepackage[mathlines]{lineno}

\usepackage{hyperref}
\hypersetup{ colorlinks,
    citecolor=blue,
    filecolor=blue,
    linkcolor=blue,
    urlcolor=blue
}

\usepackage{xcolor}

\definecolor{gray}{rgb}{0.6,0.6,0.6}
\definecolor{red}{rgb}{0.85,0,0}
\definecolor{green}{rgb}{0,0.85,0}
\definecolor{blue}{rgb}{0,0,0.85}
\definecolor{beige}{rgb}{0.92,0.87,0.78}
\usepackage[all]{hypcap}    

\usepackage[acronym]{glossaries}
\newacronym{PTV}{PTV}{planning target volume}
\newacronym{Linac}{Linac}{linear accelerator}
\newacronym{DRR}{DRR}{digitally reconstructed radiograph}
\newacronym{4DCT}{4DCT}{four-dimensional computed tomography}
\newacronym{sX-ray}{sX-ray}{simulated X-ray}
\newacronym{NMAE}{NMAE}{Normalized Mean Absolute Error}
\newacronym{SSIM}{SSIM}{Structural Similarity Index}
\newacronym{NCC}{NCC}{Normalized Cross Correlation}
\newacronym{CT}{CT}{computed tomography}
\newacronym{ETD}{ETD}{ExacTrac Dynamic}
\newacronym{HU}{HU}{Hounsfield Unit}
\newacronym{CLAHE}{CLAHE}{Contrast Limited Adaptive Histogram Equalization}
\newacronym{STD}{STD}{standard deviation}
\newacronym{FOV}{FOV}{field of view}
\newacronym{ALARA}{ALARA}{As Low As Reasonably Achievable}
\newacronym{TEM}{TEM}{tissue equivalent material}
\newacronym{IGRT}{IGRT}{image-guided radiation therapy}
\newacronym{BMI}{BMI}{body mass index}

\renewcommand{\hl}{\textcolor{black}}

\newcommand{\titlename}[0]{\sethlcolor{pink}\hl{Evaluation of attenuation-based simulated X-ray images for image-guided radiotherapy}}

\begin{document}
\doublespacing
\include{titlepage.tex}


\pagenumbering{roman}
\setcounter{page}{1}
\pagestyle{plain}

\newpage

\cen{\sf {\Large {\bfseries \titlename}}}


\begin{abstract}
\noindent {\bf Background:} In \gls{IGRT}, in-room X-rays are often used for patient positioning and monitoring, however, usually default imaging settings are utilized for each indication. Obtaining patient-specific imaging settings (independent of the treatment site) \sethlcolor{orange}\hl{before} treatment is achievable using \acrlongpl{sX-ray}, which can facilitate image quality improvements, dose reduction and \gls{IGRT} workflow optimization.\\ 
{\bf Purpose:} This study evaluated the image quality of \acrlongpl{sX-ray}, generatable on-the-fly, for thoracic, cranial and pelvic regions.\\
{\bf Methods:} \Acrlongpl{sX-ray} were created from \gls{CT} scans with indication-specific default settings (lung: \SI{110}{\kilo\volt} \SI{16}{\milli\ampere\second}, brain: \SI{90}{\kilo\volt} \SI{10}{\milli\ampere\second}, pelvis: \SI{120}{\kilo\volt} \SI{20}{\milli\ampere\second}) using a pre-release pilot software, which mimics the geometry of the room-mounted \gls{ETD} system (Brainlab SE, Germany). The basis of this algorithm was the Beer-Lambert law. Linear attenuation coefficients were approximated from the \gls{CT} hounsfield units, while incident intensities were defined by flood images recorded with the \gls{ETD}. \gls{CT} scans were included from three anthropomorphic phantoms (Lungman (Kyoto Kagaku, Japan), \gls{ETD} cranial and pelvic verification phantom (Brainlab SE, Germany)) and 57 patients (13 lung, 20 glioblastoma, 24, prostate). \sethlcolor{cyan}\hl{For evaluation, real X-rays were used as ground truth.} Three image quality metrics --- \gls{SSIM}, \gls{NCC}, \gls{NMAE} --- complemented by histogram and intensity profile comparisons were used for evaluation.\\
{\bf Results:} \glspl{SSIM} for phantom and patient average were 0.87 and 0.80 for the thoracic, 0.94 and 0.88 for the cranial and 0.85 and 0.79 for the pelvic region. Corresponding \glspl{NCC} yielded values of 0.99 and 0.89 (lung), 0.99 and 0.99 (brain) and 0.99 and 0.60 (pelvis). \glspl{NMAE} were 4.1\% and 10.7\% (lung), 2.6\% and 3.2\% (brain) and 2.6\% and 4.0\% (pelvis). Histograms of \sethlcolor{cyan}\hl{real} X-rays and \glspl{sX-ray} were slightly shifted and showed differences at the background pixel value. Intensity profiles exhibited similar trends with some differences in absolute values. \\
{\bf Conclusions:} The investigated \acrlongpl{sX-ray} demonstrated promising quality for individualized \gls{IGRT} treatment preparation. \\
\end{abstract}


\newpage
\tableofcontents
\newpage

\setlength{\baselineskip}{0.7cm}      

\pagenumbering{arabic}
\setcounter{page}{1}
\pagestyle{fancy}

\glsresetall

\section{Introduction}
\doublespacing
In \gls{IGRT}, in-room X-rays are frequently utilized for daily patient positioning and monitoring, to ensure target coverage during treatment.\cite{Infusino_Trodella_Ramella_DAngelillo_Greco_Iurato_Trodella_Nacca_Cornacchione_Mameli_2015} Typically, indication-specific default imaging settings are used for X-ray acquisition regardless of the patient dimensions and classical \glspl{DRR} serve as reference images for daily positioning.\cite{McNair_Franks_vanHerk_2022} Classical \glspl{DRR} \sethlcolor{orange}\hl{are calculated as the exponential of the line integral of the linear attenuation coefficients along each CT projection ray, but are not calibrated to the imaging energy of the simulated X‑ray system.}\cite{Russakoff_Rohlfing_Mori_Rueckert_Ho_Adler_Maurer_2005} This limits their usability for direct comparison to clinical X-rays acquired during treatment as well as pre-treatment imaging setting optimization.

In X-ray imaging varying patient dimensions require different imaging settings for optimal visualization of the anatomy.\cite{Dudhe_Mishra_Parihar_Nimodia_Kumari_2024,Antonakos_Kokkinogoulis_Giannopoulou_Efstathopoulos_2026} Patient-specific imaging optimization can therefore lead to an improvement of the \gls{IGRT} workflow by increasing the information content of the \sethlcolor{cyan}\hl{intra-treatment} X-rays, while reducing the imaging dose in accordance with the \gls{ALARA} principle. This is especially relevant for patients receiving multi-fraction treatment courses or intra-fractional imaging.\cite{Velonis_Papanastasiou_Hatziioannou_Siountas_Kamperis_Papavasileiou_Koukourakis_Seimenis_2023,Ng_Booth_Poulsen_Kuncic_Keall_2013} 
By accurately simulating the \sethlcolor{orange}\hl{imaging system}, optimized settings could be established \sethlcolor{cyan}\hl{during treatment preparation} via \glspl{sX-ray}. Such \glspl{sX-ray} \sethlcolor{cyan}\hl{provide closer resemblance to the acquired X-rays than classical} \glspl{DRR} \sethlcolor{cyan}\hl{and have therefore the potential to improve both X-ray-based patient positioning and monitoring.}

Generation of \glspl{sX-ray} \sethlcolor{orange}\hl{utilizes a variety of methods. Monte Carlo simulations demonstrate high fidelity but are too slow to be implemented in efficient clinical workflows, while recently developed machine learning approaches show promising results but require extensive training data sets. In contrast, deterministic Beer-Lambert models, that are used for DRR generation, are time-efficient but usually employ a non–energy-calibrated attenuation mapping.}\cite{Giersch_Durst_2008,Unberath_Zaech_Lee_Bier_Fotouhi_Armand_Navab_2018,Russakoff_Rohlfing_Mori_Rueckert_Ho_Adler_Maurer_2005}

The aim of this study was to evaluate the quality of deterministic, \sethlcolor{orange}\hl{spectrally calibrated} \glspl{sX-ray}, obtainable on-the-fly, for the purpose of optimizing imaging settings on a patient-individual basis (independent of the treatment site) for \gls{IGRT} \sethlcolor{orange}\hl{before treatment}. Investigated anatomical regions were lung, brain and pelvis in both phantoms and patients. The evaluation was based on \sethlcolor{cyan}\hl{real X-rays as ground truth} and three image quality metrics, as well as histogram and intensity profile comparisons.

\section{Methods}

\subsection{Background of \acrlong{sX-ray} creation}

A \sethlcolor{orange}\hl{commercial} pre-release pilot software, that is currently integrated into the \gls{ETD} v2.5 workflow (Brainlab SE, Germany) was used to generate \glspl{sX-ray}, \sethlcolor{cyan}\hl{as illustrated in Supplementary Fig. S1}.\cite{JIN2008124} The software employed \gls{CT} scans, GPU-accelerated 3D ray casting and the Beer-Lambert law

\begin{equation}
I = I_0 \exp \left( -\sum_{i=1}^{n} \mu_i \Delta x_i \right).
\label{beer_lambert_sum}
\end{equation}

In Eq. \ref{beer_lambert_sum} $I_0$ is the incident intensity of the \gls{ETD} imagers and $\mu_i$ and $\Delta x_i$ are the linear attenuation coefficient and the thickness of the $i$-th \gls{CT} layer. $I$ is the transmitted beam intensity after passing through $n$ \gls{CT} layers. 

\subsection{Calibration measurements}

\begin{figure}[htbp]
   \begin{center}
   \includegraphics[width=15cm]{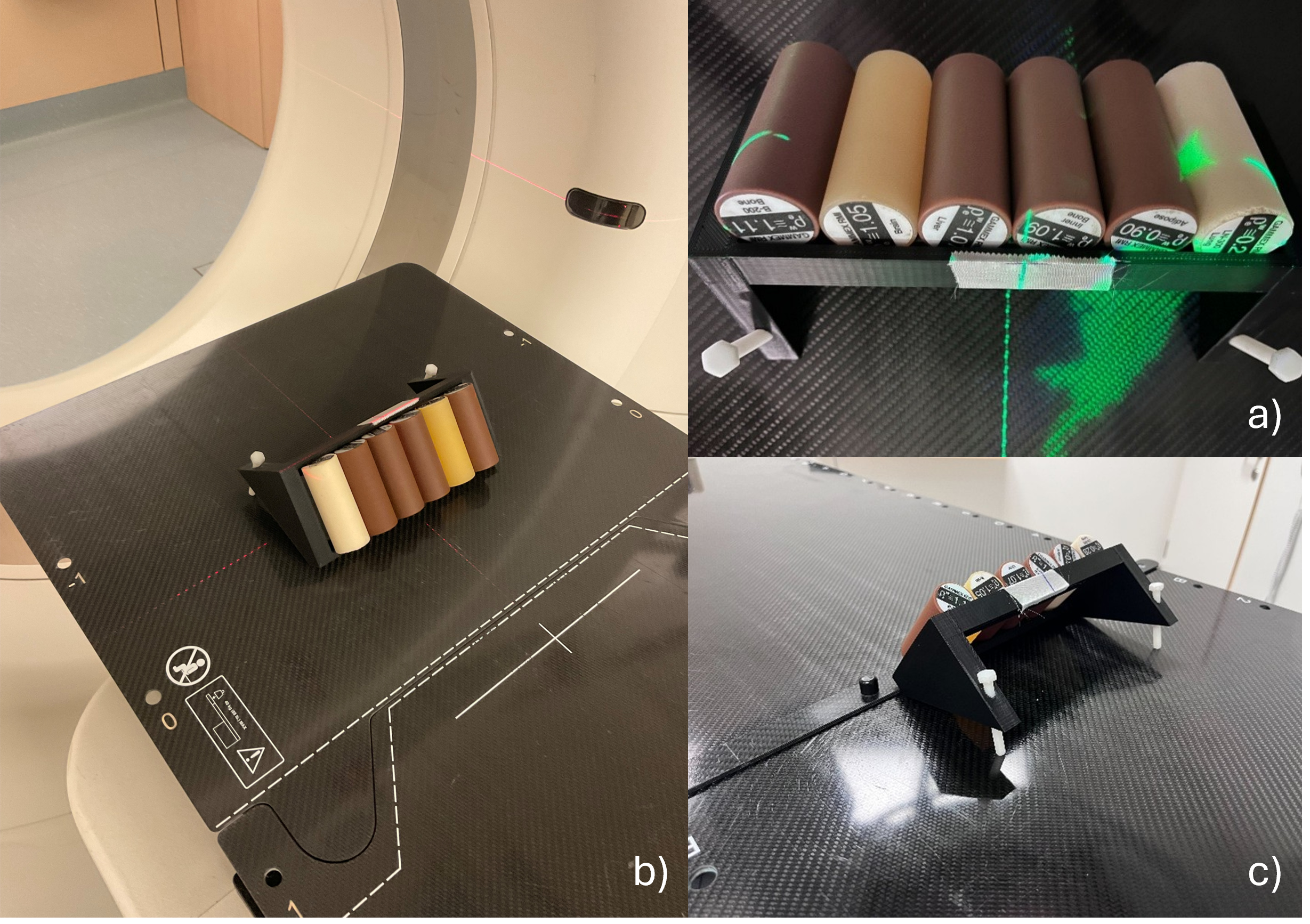}
   \captionv{12}{sX-ray calibration setup}
   {\textbf{sX-ray calibration setup:} a) Six exemplary \glspl{TEM} used for \gls{sX-ray} calibration. b) \gls{TEM} setup for determining the \gls{HU} with a \gls{CT} scanner. c) \gls{TEM} setup for determining the attenuation with an \gls{ETD} system.
   \label{fig:gammex}}  
   \end{center}
\end{figure}

\sethlcolor{cyan}\hl{Incident intensities $I_0$ were calibrated for three} \gls{ETD} \sethlcolor{cyan}\hl{systems (Brainlab SE, Germany, software version v2.0) available at our institute by acquiring flood images (X-rays without the patient couch) with different imaging settings. These partly accounted for the energy spectrum and the heel effect, though not for polychromatic beam hardening or scatter.}\cite{Kusk_Jensen_Gram_Nielsen_Precht_2021,Abdurahman_Frysch_Bismark_Melnik_Beuing_Rose_2018} The \gls{ETD} systems were mounted in irradiation rooms equipped with a Versa HD linear accelerator (Elekta AB, Sweden). Two \gls{ETD} systems had a similar imaging geometry, whereas the third differed slightly in its arrangement. \sethlcolor{cyan}\hl{Linear attenuation coefficients were determined from X-ray and} \gls{CT} \sethlcolor{cyan}\hl{image analysis of multiple} \glspl{TEM} \sethlcolor{cyan}\hl{(Gammex Inc., USA; PTW Freiburg GmbH, Germany) with known electron densities and dimensions.} Due to the oblique imaging geometry of the \gls{ETD} systems, the \glspl{TEM} were positioned on a 3D printed holder with adjustable angle to account for variations in \gls{ETD} imaging geometries. This enabled X-ray imaging along the longitudinal and transversal \gls{TEM} direction and thus a precise determination of the length of the attenuation path using multiple imaging settings. The \glspl{CT} were acquired in the same geometry and using the same holder. Exemplary setups are shown in Fig. \ref{fig:gammex}.
\sethlcolor{cyan}\hl{An assignment between} \gls{HU} \sethlcolor{cyan}\hl{and linear attenuation coefficients was established using} \gls{HU}\sethlcolor{cyan}\hl{–electron density calibration curves and was verified by} \gls{HU} analysis of the \glspl{TEM} \sethlcolor{cyan}\hl{for different CT protocols.}

\subsection{Phantom and patient data}
\label{sec:phantom_patient}
The data set included \gls{CT} scans for \glspl{sX-ray} generation and \sethlcolor{cyan}\hl{real} X-rays \sethlcolor{cyan}\hl{as ground truth} for comparison, representing three indications - lung cancer, and tumors located in the cranial and pelvic region - using both anthropomorphic phantoms and retrospective patient data. Patient \gls{CT} scans, acquired with a Siemens Somatom Definition AS scanner (Siemens Healthineers, Germany), and patient X-rays were acquired between 2023 and 2025 during clinical practice. Their retrospective use was approved by the ethics committee of the Medical University of Vienna (IRB: 1359/2025). \sethlcolor{cyan}\hl{Real} X-rays and \glspl{sX-ray} had a pixel depth of 16-bit and dimensions of 768$\times$768 pixels.

Stereoscopic patient X-rays were acquired for positioning with the \gls{ETD} system using default settings: \SI{110}{\kilo\volt}, \SI{16}{\milli\ampere\second} for the lung, \SI{90}{\kilo\volt}, \SI{10}{\milli\ampere\second} for the brain and \SI{120}{\kilo\volt}, \SI{20}{\milli\ampere\second} for the pelvis. The data set included 13 lung cancer patients (X-rays were acquired without trigger in free breathing, \gls{BMI} \sethlcolor{pink}\hl{19.8-33.8}), 20 glioblastoma patients (cranial location) and 24 prostate cancer patients (pelvic location, \gls{BMI} \sethlcolor{pink}\hl{18.1-39.9}). Available \gls{CT} scans were used for \gls{sX-ray} simulation, which all had a slice thickness of \SI{2}{\milli\meter} and a tube voltage of \SI{120}{\kilo\volt}. The applied convolution kernels for reconstruction were \textit{I30f} for the thoracic \gls{4DCT} scans, \textit{J30s} for the cranial and \textit{I30f} (in one case \textit{B30f}) for the pelvic patient \glspl{CT}.

Employed phantoms \sethlcolor{pink}\hl{(representing different anatomical dimensions)} were the Lungman anthropomorphic thorax phantom (Kyoto Kagaku, Kyoto, Japan) \sethlcolor{pink}\hl{(life-sized male thorax model,  size:} \qtyproduct{43 x 40 x 48}{\centi\metre}), and two imaging verification phantoms for the cranial \sethlcolor{pink}\hl{(life-sized skull, size:} \qtyproduct{15 x 22 x 25}{\centi\metre}) and pelvic \sethlcolor{pink}\hl{(life-sized male pelvis, size:} \qtyproduct{32 x 22 x 29}{\centi\metre}) region (Brainlab SE, Germany). \glspl{CT} with site-specific clinical protocols were acquired and had the same slice thicknesses and tube voltages as the patient scans. The applied convolution kernels for reconstruction were \textit{I31f}, \textit{J30s} and \textit{I30f} for the Lungman, cranial and pelvic phantom, respectively. \sethlcolor{pink}\hl{For the cranial phantom an additional} \gls{CT} \sethlcolor{pink}\hl{was acquired with a slice thickness of} \SI{1}{\milli\meter} for comparison \sethlcolor{pink}\hl{(see Supplementary Chapter B)}. Stereoscopic \gls{ETD} X-rays of the phantoms were recorded with the default settings.

\subsection{\Acrlong{sX-ray} creation and evaluation}
\label{sec:eval}
\sethlcolor{cyan}\hl{Real stereoscopic} X-rays were fused with spectrally uncalibrated \glspl{sX-ray} (thoracic and cranial region) or \glspl{DRR} (pelvis) generated from the CT scan using Brainlab 6D fusion to determine patients positioning shifts between \gls{CT} and \sethlcolor{cyan}\hl{real} X-rays.\cite{JIN2008124} This was necessary to align the two projection images and to account for anatomical changes before creating the spectrally calibrated \glspl{sX-ray}. For the \glspl{4DCT} of the lung cancer patients, the \gls{CT} phase with the best fusion result measured by means of cross-correlation was chosen for evaluation, since there was no ground truth for the breathing phase of the \sethlcolor{cyan}\hl{real} X-ray. Input parameters for \gls{sX-ray} generation were the \gls{CT} with its isocenter coordinate, the 6D fusion shift (to minimize offset errors between \sethlcolor{cyan}\hl{real} X-rays and \glspl{sX-ray}), the imaging setting (same imaging setting as used for the corresponding \sethlcolor{cyan}\hl{real} X-ray) and the incident intensity $I_0$ and imaging geometry file of the respective \gls{ETD} system. Since a pixel value crop at an intensity value of 50000 \sethlcolor{cyan}\hl{(maximum intensity)} was applied to the clinical \gls{ETD} X-rays by default (hardcoded in the \gls{ETD} system) to overcome technical limitations, this was also applied to the \glspl{sX-ray}.
\sethlcolor{cyan}\hl{Real} X-ray and \gls{sX-ray} images displayed in this manuscript were filtered with \gls{CLAHE} (python module openCV version 4.11.0: clipLimit=50, tileGridSize=[16$\times$16]) and gamma corrected ($\gamma$ = 1.5) for visualization purposes only.

Generated raw \gls{sX-ray} images were compared to corresponding \sethlcolor{cyan}\hl{real} X-rays without applying a post-processing filter. \sethlcolor{pink}\hl{One part of the analysis comprised} three different image quality metrics: the \gls{SSIM} for measuring perceptual similarity (python module openCV version 4.11.0), the \gls{NCC} for evaluating alignment (python module scikit-image version 0.25.2) and the \gls{NMAE} for calculating the relative pixel value differences in \% (normalization to pixel value range).\cite{wang2004image,lewis1995fast,willmott2005advantages} For the patient images, the whole image and the \gls{PTV} region only \sethlcolor{cyan}\hl{(3D contours projected onto 2D images)} were evaluated separately. Phantom images were always compared without region focus, evaluating the entire image. An average of the two \gls{ETD} tubes was considered for the metrics and averages ($\pm$ \gls{STD}) over all patients were calculated and reported for each indication. \sethlcolor{pink}\hl{Another part of the analysis focused on} pixel value histograms averaged over all patients for the \sethlcolor{cyan}\hl{real} X-rays and \glspl{sX-ray}. Histograms were compared by evaluating peak pixel value positions and their corresponding counts. Differences were reported as percentage of the total image pixel count and the entire pixel value range. \sethlcolor{pink}\hl{Lastly, the analysis addressed} horizontal intensity profiles through the center of the image of one of the two \gls{ETD} tubes for the phantoms and one exemplary patient per indication.

\subsection{Influence of different incident intensities and imaging settings}

The impact of different incident intensities on \gls{sX-ray} image quality metrics was assessed using the Lungman phantom, since it was the most anatomically detailed among the three phantoms. Generation of \glspl{sX-ray} was performed with $I_0$ values from the three institutional \gls{ETD} systems, including the actual system used for the \sethlcolor{cyan}\hl{real} X-ray acquisition. 

To assess the impact of different tube voltages and currents, X-rays of all three phantoms were additionally acquired in the following imaging setting range: \SIrange{70}{130}{\kilo\volt}, \SIrange{4}{32}{\milli\ampere\second}. \sethlcolor{cyan}\hl{Real} X-ray and \gls{sX-ray} pixel value histograms averaged over all imaging settings were analyzed for all phantoms as described in Sec. \ref{sec:eval}. The \gls{sX-ray} image quality metric dependence on imaging setting was assessed using the Lungman phantom only.

\section{Results}

The incident intensities $I_{0}$ for the two \gls{ETD} systems with similar geometry differed approximately by 3\%, while the slightly different system showed a deviation of roughly 23\% from the other two.

\gls{sX-ray} image quality metrics were nearly always superior for the phantoms compared to their average patient-derived counterparts (Tab. \ref{tab:metrics} and Supplementary Fig. S2). Without focus region, the \gls{SSIM} was higher for phantoms by 8.8\%, 6.8\% and 7.6\% for lung, brain and pelvis while the \gls{NMAE} was lower by 61.7\%, 18.8\% and 35.0\%. 
The \gls{NCC} showed an increase for lung \sethlcolor{pink}\hl{and} pelvis only (11.2\% and 65.0\%) and no difference for the cranial region. Overall, the cranial region experienced the smallest decline in image quality when transitioning from phantom to patient X-rays, due to the rigid anatomy. With focus region, the \gls{SSIM} for the phantoms was higher by 6.1\%, 17.5\% and 4.9\%, and by 2.1\%, 2.1\% and 5.3\%  for the \gls{NCC} in the thoracic, cranial and pelvic region, respectively. The \gls{NMAE} for the phantoms was lower by 31.7\% for the thoracic and higher by 52.9\% for the pelvic region. No change was observed in the brain. These results showed that, especially for the lung cohort, the reduction of the evaluation region and focus on the primary area of clinical interest was essential.

\begin{table}[htbp]
\centering
\label{tab_label}
\vspace{2ex}
\setlength{\tabcolsep}{3.5pt} 
\small 
\begin{tabular}{|l|ccc|ccc|ccc|}
\hline
 & \multicolumn{3}{c|}{Phantom} & \multicolumn{3}{c|}{\makecell{Patient mean\\$\pm$STD}} & \multicolumn{3}{c|}{\makecell{Patient mean\\$\pm$STD (\gls{PTV})}} \\
\cline{2-10}
Indication & SSIM & NCC & NMAE (\%) & SSIM & NCC & NMAE (\%) & SSIM & NCC & NMAE (\%)\\
\hline
Lung     & 0.87 & 0.99 & 4.1 & \makecell{0.80  \\ $\pm$0.05}& \makecell{0.89 \\ $\pm$0.07} & \makecell{10.7 \\ $\pm$6.4} & \makecell{0.82 \\ $\pm$0.12} & \makecell{0.97 \\ $\pm$0.03} & \makecell{6.0 \\ $\pm$6.0} \\
Brain     & 0.94 & 0.99 & 2.6 & \makecell{0.88 \\ $\pm$0.02} & \makecell{0.99 \\ $\pm$0.02}  & \makecell{3.2\\ $\pm$1.5}  & \makecell{0.80\\ $\pm$0.06}  & \makecell{0.97\\ $\pm$0.01}  & \makecell{2.6\\ $\pm$0.5}  \\
Pelvis & 0.85 & 0.99 & 2.6 & \makecell{0.79\\ $\pm$0.08} & \makecell{0.60\\ $\pm$0.17} & \makecell{4.0\\ $\pm$2.2} & \makecell{0.81\\ $\pm$0.14} & \makecell{0.94\\ $\pm$0.04} & \makecell{1.7\\ $\pm$0.7} \\
\hline
\end{tabular}
\caption{\textbf{Image quality metrics:} Image quality metrics for phantoms and patients (whole image vs. \gls{PTV} only) at default settings. The patient image quality metrics represent an average $\pm$ \gls{STD} over all patients of the respective cohort for the two evaluation regions.
\label{tab:metrics}}
\end{table}

\begin{figure}[htbp]
   \begin{center}
   \includegraphics[width=15cm]{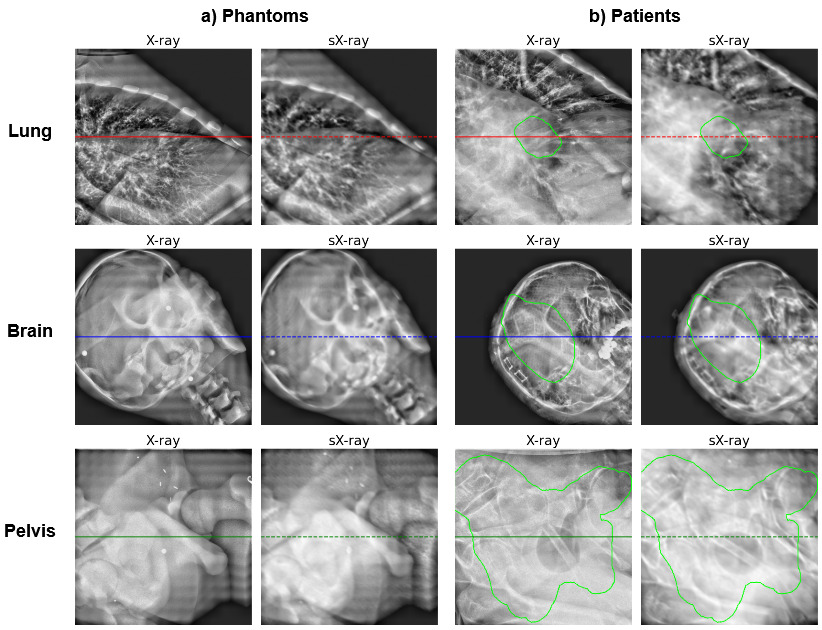}
   \captionv{12}{X-rays and \acrlongpl{sX-ray}}
   {\textbf{X-rays and \glspl{sX-ray}:} \sethlcolor{cyan}\hl{Real} X-rays and corresponding \glspl{sX-ray} for the phantoms in a) and an exemplary patient for each indication in b) with indicated \gls{PTV} in light green (top: lung; middle: brain; bottom: pelvis). Lines for intensity profile comparison (Fig. \ref{fig:intensity_profiles}) are indicated as well. All images were produced with the respective default setting. 
   \label{fig:xray_images}}  
   \end{center}
\end{figure}

For illustrative purposes, Fig. \ref{fig:xray_images} shows \sethlcolor{cyan}\hl{real} X-ray and \gls{sX-ray} images for all phantoms and an exemplary patient for each indication.

\begin{figure}[htbp]
   \begin{center}
   \includegraphics[width=17cm]{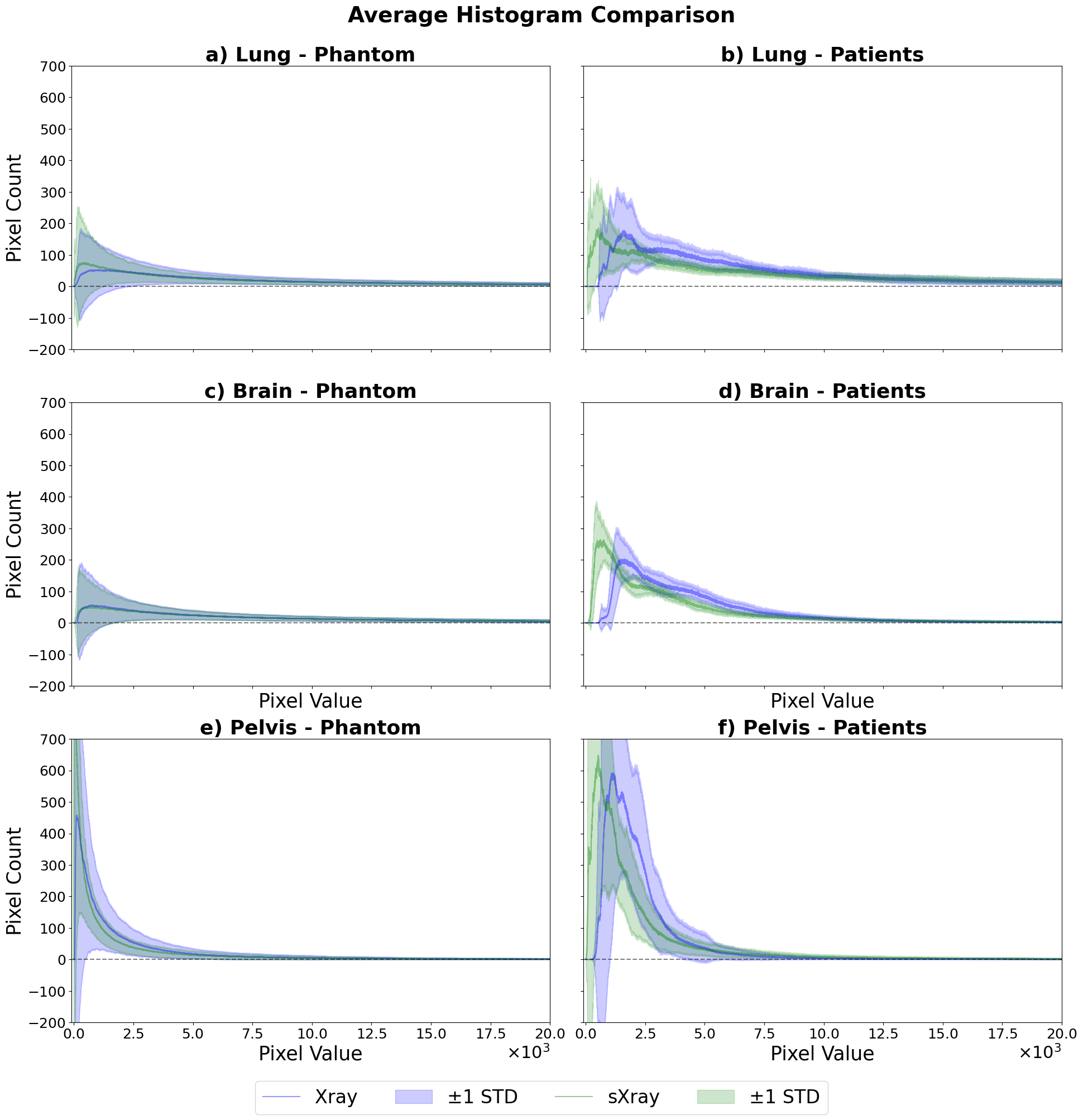}
    \captionv{12}{Histogram comparison}{\textbf{Histogram comparison:} a), c) and e) show average pixel counts over all imaging settings for the phantoms. b), d) and f) show corresponding patient averages considering all patients of the respective cohort.
    \label{fig:histo_differences_lung_cranial_prostate}}
    \end{center}
\end{figure}

Fig. \ref{fig:histo_differences_lung_cranial_prostate} shows average histogram comparisons for all imaging settings used for the phantoms and all patients, respectively. Compared to the average \sethlcolor{cyan}\hl{real} X-ray histograms, the peak positions of the average \gls{sX-ray} histograms (excluding the background pixel value) generally shifted toward lower values, except for the cranial phantom (shift by 0.1\% to higher values). For the Lungman phantom and the pelvic phantom the peak shifts were 1.3\% and 0.2\% (pelvis). For the corresponding patients, where only the respective default imaging setting was available (no averaging over different imaging settings), the peak shifts were larger: 2.1\% (lung), 2.1\% (brain) and 1.2\% (pelvis). All percentages are given relative to the entire pixel value range. The count difference between the maxima was less than 0.2\% with respect to the total image pixel count of 768$\times$768=589824 pixels in all cases. The background pixel value of 50000 (maximum intensity) showed histogram deviations as well, corresponding to differences in the interface between the subject (phantom or patient) and the background. The average relative pixel value count differences between \glspl{sX-ray} and \sethlcolor{cyan}\hl{real} X-rays at the background pixel value 50000 were roughly -0.8\% (Lungman phantom), 11.4\% (lung cancer patients), 0.7\% (cranial phantom), -1.1\% (glioblastoma patients), -0.1\% (pelvic phantom) and 1.3\% (prostate cancer patients) of the total image pixel count.

\begin{figure}[htbp]
    \centering
    \includegraphics[width=17cm]{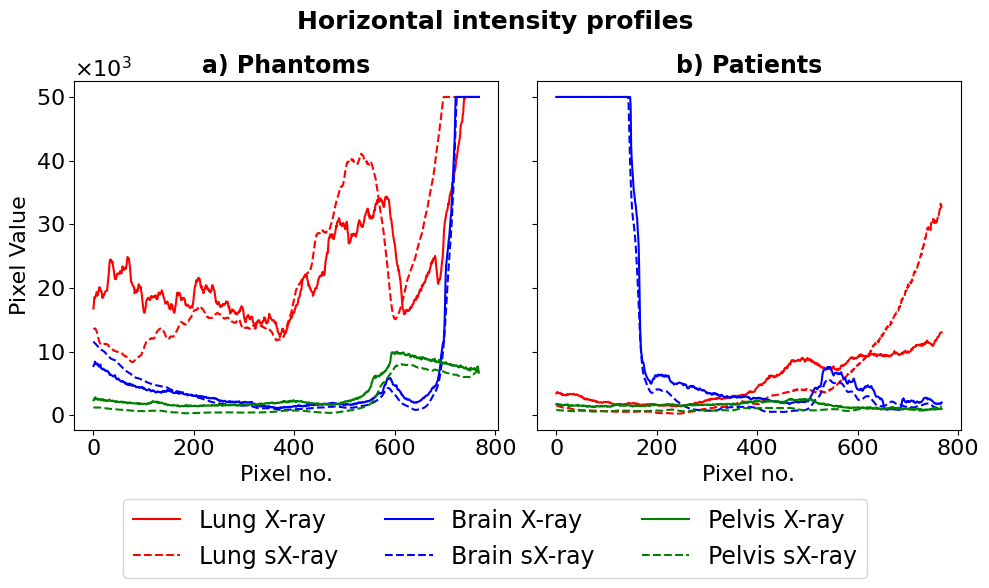}
    \captionv{12}{Intensity profile comparison}{\textbf{Intensity profile comparison:} Intensity profiles along horizontal lines trough \sethlcolor{cyan}\hl{real} X-ray and \gls{sX-ray} images of a) the phantoms and b) an exemplary patient for each indication. Images were produced with the respective default setting and are shown in Fig. \ref{fig:xray_images} with indicated lines.
    \label{fig:intensity_profiles}}
\end{figure}

Fig. \ref{fig:intensity_profiles} shows intensity profiles along the horizontal lines indicated in  Fig. \ref{fig:xray_images}. While \glspl{sX-ray} and \sethlcolor{cyan}\hl{real} X-rays exhibited similar trends, noticeable intensity offsets persisted. Intensity profiles for brain and pelvis generally demonstrated higher fidelity than for lung.

Using alternative incident intensities $I_0$ for \gls{sX-ray} generation of the Lungman phantom generally showed a decrease in image quality metrics. The \gls{SSIM} and \gls{NCC} decreased by 1.0\% and 0.8\% for the systems with similar $I_0$, while 7.7\% and 0.9\% for the system with deviating $I_0$. The \gls{NMAE} increased by 20.6\% and 131.5\%, respectively.

\begin{figure}[H]
   \begin{center}
   \includegraphics[width=16cm]{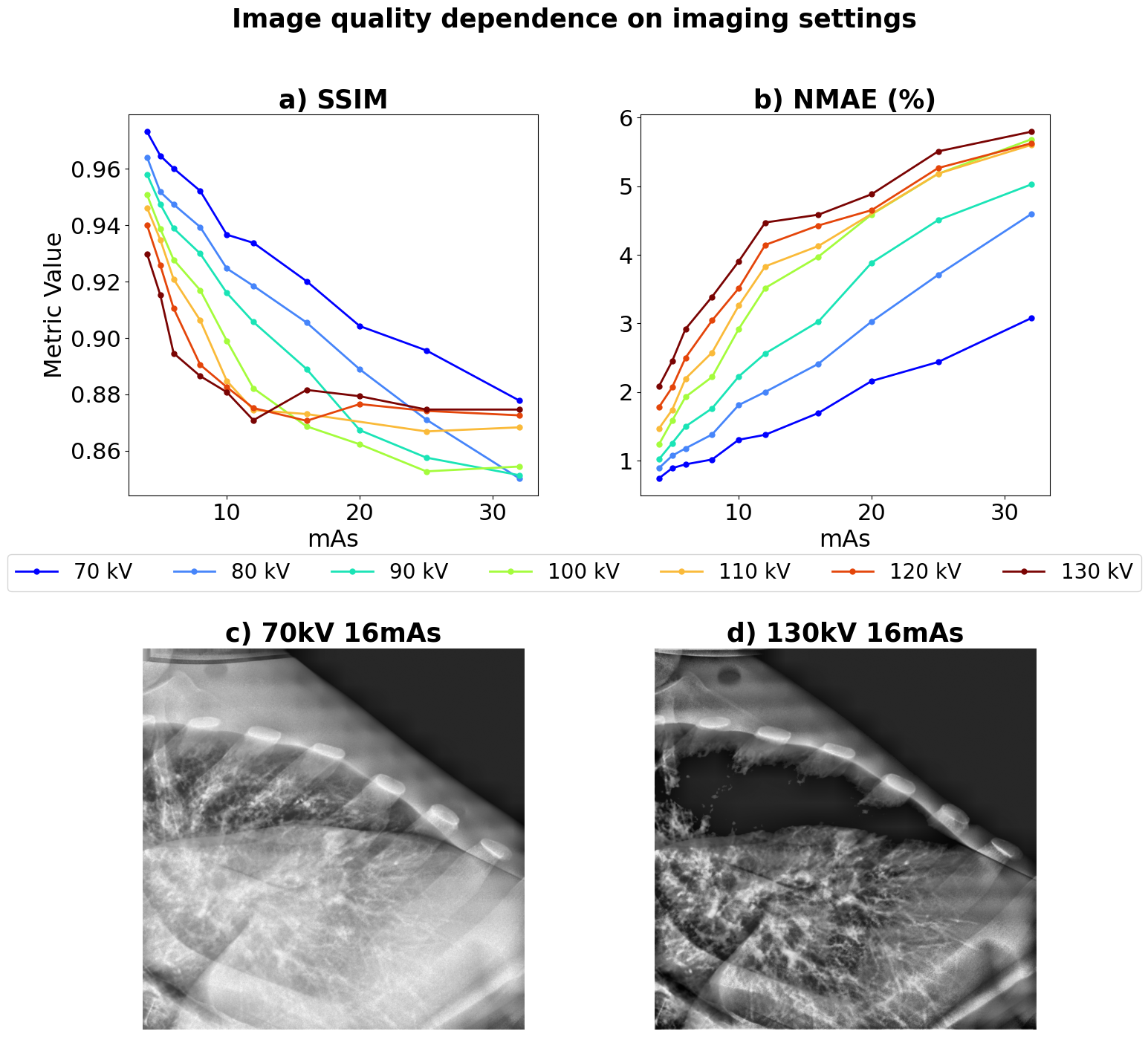}
   \captionv{12}{Image quality metric dependence on imaging settings}
   {\textbf{Image quality metric dependence on imaging settings:} a) SSIM and b) NMAE dependence on imaging settings for the Lungman phantom. The \gls{NCC} change was not plotted, since it was negligible. c) and d) show \sethlcolor{cyan}\hl{real} X-rays acquired with non-default settings for visualization.
   \label{fig_metric_dependence} 
    }  
    \end{center}
\end{figure}

Among varying imaging settings, higher tube voltages and currents generally increased the difference between \sethlcolor{cyan}\hl{real} X-rays and \glspl{sX-ray}. The \gls{SSIM} decreased up to 12.6\% and the \gls{NMAE} increased up to 683.0\%, as shown in Fig. \ref{fig_metric_dependence}. The \gls{NCC} decrease was less than 0.2\%. In general, for higher imaging energies, more saturation effects \sethlcolor{cyan}\hl{(maximum measurable intensity of the detector at a pixel value of 50000)} occurred, especially in the lung region (Fig. \ref{fig_metric_dependence}d).

\section{Discussion}

The investigated \glspl{sX-ray}, which are obtainable on-the-fly, provided satisfactory image quality \sethlcolor{pink}\hl{for different indications and patient sizes} and can serve as basis to optimize X-ray settings on a patient-specific basis and by that improve clinical \gls{IGRT} workflows. For phantom images, \gls{SSIM} and \gls{NCC} were greater than 0.85 and 0.99, respectively, when using indication-specific default imaging settings. The \gls{NMAE} was at most 4.13\%, demonstrating good agreement between \sethlcolor{cyan}\hl{real} X-rays and \glspl{sX-ray}. For the patient data, the metrics deteriorated, but were larger than 0.79 and 0.60 for \gls{SSIM} and \gls{NCC} and at most 10.70\% for \gls{NMAE}, when evaluated across the whole image. Comparison of \gls{sX-ray} and \sethlcolor{cyan}\hl{real} X-ray histograms showed a slight shift and differences in pixel value counts in the background region, representing imprecise phantom/patient boundaries. 

The evaluation utilized a 6D fusion of \sethlcolor{cyan}\hl{real} X-rays and \glspl{sX-ray}, a process susceptible to uncertainties arising from anatomical or positional variations between CT and X-ray acquisition, especially relevant for the patients.\cite{JIN2008124,Infusino_Trodella_Ramella_DAngelillo_Greco_Iurato_Trodella_Nacca_Cornacchione_Mameli_2015} Anatomical differences were particularly evident in the prostate cancer cohort (see for example position of the rectal balloon in Fig. \ref{fig:xray_images}b), resulting in \sethlcolor{orange}\hl{a wide spread of} \gls{NCC} values with a modest average of 0.60, when evaluated across the whole image. Conversely, these discrepancies were less pronounced in the lung cancer cohort and even less so in the glioblastoma cases, where structural consistency was higher. A higher \gls{NCC} for the prostate cancer patients was achieved for metric evaluation in the \gls{PTV} region only. For the phantoms, imperfections in the 6D fusion (in addition to simulation limitations) likely also contributed to discrepancies between \sethlcolor{cyan}\hl{real} X-rays and \glspl{sX-ray}, \sethlcolor{pink}\hl{especially for the lung phantom, featuring the highest anatomical complexity (see discrepancies in Fig.} \ref{fig:intensity_profiles}a).

Histogram analysis revealed two main differences between \glspl{sX-ray} and \sethlcolor{cyan}\hl{real} X-rays. The first was a slight histogram shift, especially evident in the patient cases, where only one imaging setting was available. Averaging over several imaging settings for the phantoms reduced the peak shifts, indicating good overall histogram agreement. An improved simulation of beam hardening might reduce the current underestimation of intensity in the \glspl{sX-ray}.\cite{Abdurahman_Frysch_Bismark_Melnik_Beuing_Rose_2018,Unberath_Zaech_Gao_Bier_Goldmann_Lee_Fotouhi_Taylor_Armand_Navab_2019} The second noticeable difference occurred at the pixel value 50000, which corresponds to the maximum intensity (background). This suggests imprecise simulation of the transition between phantom/patient and background or slight fusion errors. These deviations are also possible in regions containing saturation effects in the subject region, \ie the same intensity as in the background. However, the observed differences at the background value were less than 2\% of the total image pixel count using default settings (excluding the lung cancer cohort), which suggests that subject-background boundary simulation works sufficiently.

In some patient cases, the \gls{CT} datasets exhibited a limited longitudinal \gls{FOV}, leading to anatomical truncation in the \gls{sX-ray} projections relative to the clinically acquired X-ray images. This was most evident for the lung cancer patient cohort and can be observed in Fig. \ref{fig:xray_images}b and the high average relative pixel value count differences at the background value between \sethlcolor{cyan}\hl{real} X-rays and \glspl{sX-ray} of 11.4\%, which can not be solely attributed to imprecise simulation. While whole image metrics were negatively impacted by these \gls{FOV} discrepancies, \gls{PTV} only \sethlcolor{orange}\hl{image quality} metrics remained unaffected (see e.g., \gls{NCC} \sethlcolor{orange}\hl{for the lung cancer patients)}, due to the central location within the planning \gls{CT}.

Using $I_{0}$ values from different \gls{ETD} systems led to a considerable deterioration in metrics, especially the \gls{NMAE}. Therefore, it is recommendable to calibrate initial intensities for each \gls{ETD} system separately to minimize errors in the \gls{sX-ray} simulation procedure.

\sethlcolor{pink}\hl{Unmodeled effects, such as scatter and full detector response, likely propelled the overall degradation in simulated image quality at higher energies, while detector saturation and energy-dependent beam hardening could explain the subsequent plateaus and their differing levels} for the \gls{SSIM}. However, the correspondence between the \gls{sX-ray} and \sethlcolor{cyan}\hl{real} X-ray images was still acceptable when increasing the imaging energy (Fig. \ref{fig_metric_dependence}), which demonstrates applicability for setting optimization aimed at making relevant anatomical structures visible in the X-rays.

In general, the sharpness of the \glspl{sX-ray} was limited by the resolution of the \gls{CT}, which was \SI{2}{\milli\meter} in this study corresponding to clinical practice.\cite{Ahmed_Garzanich_Melaragno_Nyirjesy_Windheim_Marquardt_Luttrull_Quails_VanKoevering_2024} Increasing the \gls{CT} resolution could improve the \gls{sX-ray} quality further, \sethlcolor{pink}\hl{as shown in the Supplementary Chapter B. For an envisaged improvement of the treatment preparation workflow, standard protocols might, however, be sufficient.}

For comparison, Pointon et al.\cite{Pointon_Wen_Tugwell-Allsup_Sujar_Letang_Vidal_2023} reported a \gls{SSIM} between a deterministically simulated X-ray and corresponding \sethlcolor{cyan}\hl{real} X-ray of 0.94 for the Lungman phantom, while we obtained an \gls{SSIM} value of 0.87. However, a direct comparison is challenging, since the \gls{CT} (resolution and protocol) and X-ray (imaging system and setting) data was not identical.

The explored \glspl{sX-ray} could facilitate patient-specific optimization of imaging settings \sethlcolor{orange}\hl{before X-ray based} \gls{IGRT} and partly replace classical \glspl{DRR} in registration algorithms \sethlcolor{orange}\hl{for online treatment monitoring}. In addition, \glspl{sX-ray} could serve as an educational tool for training personnel in establishing meaningful imaging settings while gaining familiarity with the expected visual presentation or support radiotherapy through data augmentation for deep learning models.\cite{Huang_Kurz_Freislederer_Manapov_Corradini_Niyazi_Belka_Landry_Riboldi_2024, Pointon_Wen_Tugwell-Allsup_Sujar_Letang_Vidal_2023} However, despite high image quality metrics and good alignment of histograms and intensity profiles, noticeable discrepancies remained. Advancing to highly sensitive applications, such as radiologist training, will likely necessitate more sophisticated simulation techniques to achieve closer physical resemblance.\cite{Koetzier_Wu_Mastrodicasa_Lutz_Chung_Koszek_Pratap_Chaudhari_Rajpurkar_Lungren_et_al._2024}

\section{Conclusions}

The investigated \glspl{sX-ray} showed clinically acceptable accuracy for individualized \gls{IGRT} treatment preparation. Increasing the \gls{CT} resolution or \sethlcolor{orange}\hl{refining the simulation method} could further improve the resemblance between \glspl{sX-ray} and \sethlcolor{cyan}\hl{real} X-rays.

\newpage

\clearpage

\addcontentsline{toc}{section}{\numberline{}Appendix}









\section*{References}
\addcontentsline{toc}{section}{\numberline{}References}


\bibliographystyle{./medphy.bst}    


\clearpage
\pagestyle{empty}
\nolinenumbers

\setcounter{section}{0} 
\renewcommand{\thesubsection}{\Alph{subsection}}
\section*{Supplementary material}

\setcounter{figure}{0} 
\renewcommand{\thefigure}{S\arabic{figure}}
\setcounter{table}{0} 
\renewcommand{\thetable}{S\arabic{table}}

\subsection{Supporting figures}

\begin{figure}[htbp]
   \begin{center}
   \includegraphics[width=17cm]{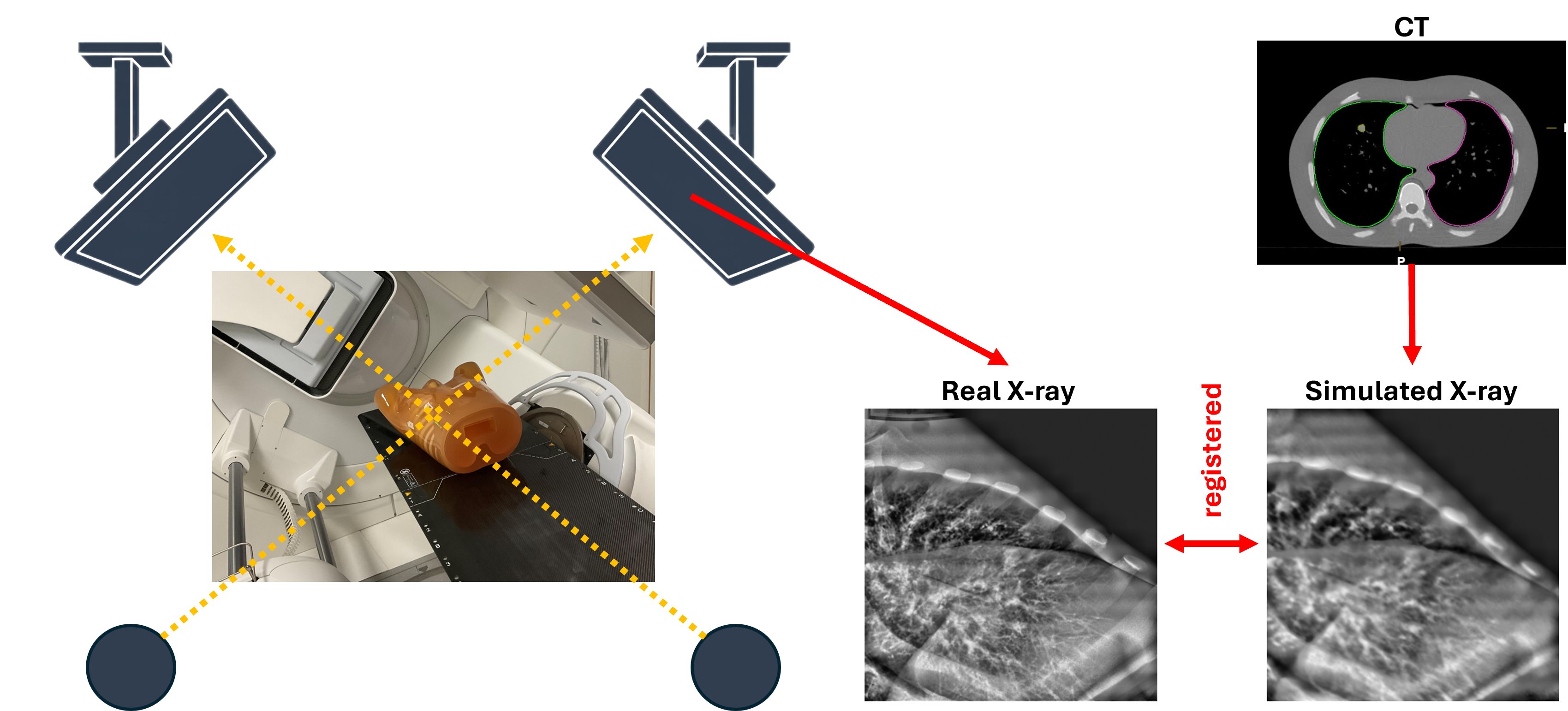}
   \captionv{12}{X-ray acquisition / simulation sketch}
   {\textbf{X-ray acquisition / simulation sketch:} Schematic sketch of X-ray acquisition with the Lungman phantom and the \gls{ETD} system. X-ray simulations from the \gls{CT} scan used the same imaging geometry.
   \label{fig_metric_boxplot} 
    }  
    \end{center}
\end{figure}

\begin{figure}[htbp]
   \begin{center}
   \includegraphics[width=17cm]{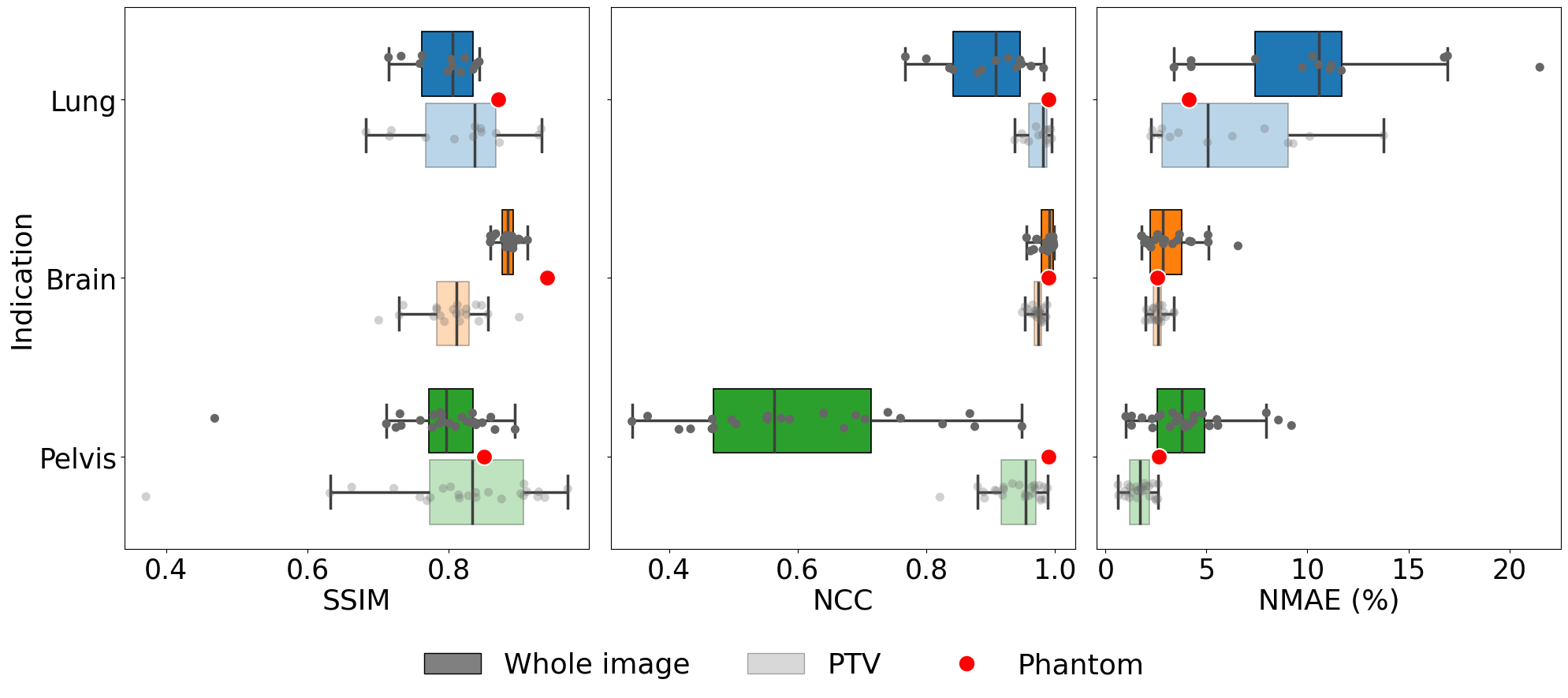}
   \captionv{12}{Image quality metric boxplots}
   {\textbf{Image quality metric boxplots:} SSIM, NCC and NMAE boxplots for all patients (grey points). Two boxplots are shown each - lung cancer (blue), glioblastoma (orange), prostate cancer (green) - representing image quality metric evaluation in the whole image (opaque) and \gls{PTV} region (translucent), respectively. The red points display the corresponding phantom metric using the default setting.
   \label{fig_metric_boxplot} 
    }  
    \end{center}
\end{figure}

\newpage

\subsection{Improved \gls{CT} resolution}

\begin{figure}[htbp]
    \centering
    \begin{subfigure}[b]{0.32\textwidth}
        \centering
        \includegraphics[width=\linewidth]{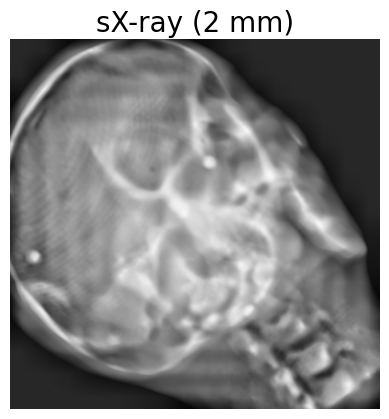}
        \caption{}
        \label{fig:2mm}
    \end{subfigure}
    \hfill
    \begin{subfigure}[b]{0.32\textwidth}
        \centering
        \includegraphics[width=\linewidth]{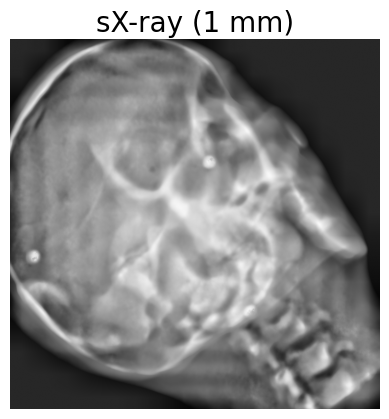}
        \caption{}
        \label{fig:1mm}
    \end{subfigure}
    \hfill
    \begin{subfigure}[b]{0.32\textwidth}
        \centering
        \includegraphics[width=\linewidth]{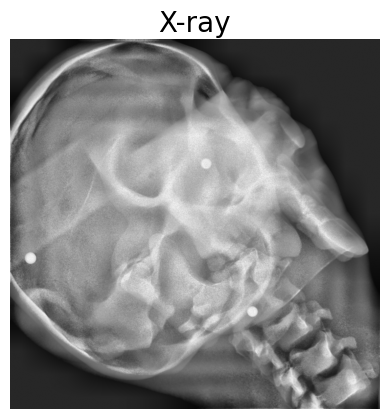}
        \caption{}
        \label{fig:1mm}
    \end{subfigure}
    \caption{\textbf{Slice thickness comparison:} \glspl{sX-ray} produced from a \gls{CT} with a) a slice thickness of \SI{2}{\milli\meter} (clinical standard) and b) a slice thickness of \SI{1}{\milli\meter}. c) shows the corresponding real X-ray.}
    \label{fig:slice_thickness}
\end{figure}

Using default X-ray imaging settings and decreasing the \gls{CT} slice thickness for the cranial phantom from \SI{2}{\milli\meter} to \SI{1}{\milli\meter} increased the \gls{SSIM} and \gls{NCC} by less than 1\%, while the \gls{NMAE} decreased by 3.1\%. Artefacts were visibly reduced, as shown in Figure \ref{fig:slice_thickness}.
\\

\end{document}

%% file: titlepage.tex
\cen{Titlepage\\
\sf {\Large {\bfseries \titlename} \\  
\vspace*{10mm} 
Hannah~Jungreuthmayer$^{1,2}$,
Julius~Arnold$^{1,2}$,
Andreas~Renner$^{1,2}$,
Wolfgang~Lechner$^{1,2}$,
Maximilian~Schmid$^{1,2}$,
Richard~Wohlgemuth$^{3}$, 
Kajetan~Berlinger$^{3}$,
Dietmar~Georg$^{1,2}$, 
Barbara~Knäusl$^{1,2}$} \\
$^{1}$Department of Radiation Oncology, Comprehensive Cancer Center Vienna, Medical University of Vienna, Vienna, Austria \\
$^{2}$Christian Doppler Laboratory for Image and Knowledge Driven Precision Radiation  Oncology, Medical University of Vienna, Austria \\
$^{3}$Brainlab SE, Munich, Germany
}
\subsubsection*{Author Contribution Statement}
{\small
\textbf{Hannah Jungreuthmayer:} Methodology, Investigation, Software, Writing - Original Draft, Writing - Review \& Editing, Visualization. \textbf{Julius Arnold:} Methodology, Investigation, Writing - Review \& Editing. \textbf{Andreas Renner:} Conceptualization, Investigation, Writing - Review \& Editing. \textbf{Wolfgang Lechner:} Conceptualization, Writing - Original Draft, Writing - Review \& Editing, Supervision. \textbf{Maximilian Schmid:}  Writing - Review \& Editing, Funding acquisition. \textbf{Richard Wohlgemuth:} Software, Resources, Writing - Review \& Editing. \textbf{Kajetan Berlinger:} Software, Resources, Writing - Review \& Editing. \textbf{Dietmar Georg:} Resources, Writing - Review \& Editing. \textbf{Barbara Knäusl:} Conceptualization, Writing - Original Draft, Writing - Review \& Editing, Supervision, Project administration, Funding acquisition.}

\subsubsection*{Ethics and IRB compliance statement}
{\small
The manuscript involved the analysis of retrospectively collected X-rays and computed tomography scans. All data utilized in this research was obtained in accordance with applicable legal and ethical requirements. This retrospective study was approved by the Ethics Committee of the Medical University of Vienna (approval: 1359/2025). We confirm that the research was conducted in accordance with the Declaration of Helsinki. Since the data was anonymized and did not involve any identifiable human subjects, explicit consent for publication was not required. No animal experimentation was involved in this study.

\subsubsection*{Acknowledgement Statement}
{\small
The financial support by the Austrian Federal Ministry of Economy, Energy and Tourism, the National Foundation for Research, Technology and Development and the Christian Doppler Research Association is gratefully acknowledged. The authors thank the company Brainlab SE for their technical support and the pre-release provision of the \acrlong{sX-ray} tool.}

\subsubsection*{Conflict of Interest Statement}
{\small
Within the frame of the Christian Doppler Research Association this study was financially supported by Brainlab SE. Richard Wohlgemuth and Kajetan Berlinger are employed by Brainlab SE.}

\subsubsection*{Data Availability Statement}
{\small
The patient data and the software tool evaluated in this study are not publicly available due to privacy restrictions and commercial confidentiality.}

\begin{flushleft}
\textbf{Corresponding author:} Hannah Jungreuthmayer\\
\textbf{Contact:} hannah.jungreuthmayer@meduniwien.ac.at 
\end{flushleft}

%% file: latex_template_MedPhys_2021.bbl
\vspace*{-15mm}

\begin{thebibliography}{10}

\bibitem{Infusino_Trodella_Ramella_DAngelillo_Greco_Iurato_Trodella_Nacca_Cornacchione_Mameli_2015}
E.~Infusino, L.~Trodella, S.~Ramella, et al.,
\newblock Estimation of patient setup uncertainty using BrainLAB Exatrac
  X‐Ray 6D system in image‐guided radiotherapy,
\newblock Journal of Applied Clinical Medical Physics {\bf 16}, 99–107
  (2015),
\newblock doi:10.1120/jacmp.v16i2.5102.

\bibitem{McNair_Franks_vanHerk_2022}
H.~A. McNair, K.~N. Franks, and M.~van Herk,
\newblock Editorial: On Target 2: Updated Guidance for Image-guided Radiotherapy,
\newblock Clinical Oncology {\bf 34}(3), 187-188
  (2022),
\newblock doi:10.1016/j.clon.2021.10.002.

\bibitem{Russakoff_Rohlfing_Mori_Rueckert_Ho_Adler_Maurer_2005}
D.~Russakoff, T.~Rohlfing, K.~Mori, et al.,
\newblock Fast generation of digitally reconstructed radiographs using
  attenuation fields with application to 2D-3D image registration,
\newblock IEEE Transactions on Medical Imaging {\bf 24}, 1441–1454 (2005),
\newblock doi:10.1109/TMI.2005.856749.

\bibitem{Dudhe_Mishra_Parihar_Nimodia_Kumari_2024}
S.~S. Dudhe, G.~Mishra, P.~Parihar, D.~Nimodia, and A.~Kumari,
\newblock Radiation Dose Optimization in Radiology: A Comprehensive Review of
  Safeguarding Patients and Preserving Image Fidelity,
\newblock Cureus  (2024),
\newblock doi:10.7759/cureus.60846.

\bibitem{Antonakos_Kokkinogoulis_Giannopoulou_Efstathopoulos_2026}
I.~Antonakos, K.~Kokkinogoulis, M.~Giannopoulou, and E.~P. Efstathopoulos,
\newblock Chest Radiography Optimization: Identifying the Optimal kV for Image
  Quality in a Phantom Study,
\newblock  {\bf 12}, 49 (2026),
\newblock doi:10.3390/jimaging12010049.

\bibitem{Velonis_Papanastasiou_Hatziioannou_Siountas_Kamperis_Papavasileiou_Koukourakis_Seimenis_2023}
M.~Velonis, E.~Papanastasiou, K.~Hatziioannou, et al.,
\newblock Dose optimization of 2D X-ray image acquisition protocols in
  image-guided radiotherapy,
\newblock Physica Medica {\bf 115}, 103161 (2023),
\newblock doi:10.1016/j.ejmp.2023.103161.

\bibitem{Ng_Booth_Poulsen_Kuncic_Keall_2013}
J.~A. Ng, J.~Booth, P.~Poulsen, Z.~Kuncic, and P.~J. Keall,
\newblock Estimation of effective imaging dose for kilovoltage intratreatment monitoring of the prostate position during cancer radiotherapy,
\newblock Physics in Medicine and Biology {\bf 58}(17), 5983-5996
  (2013),
\newblock doi:10.1088/0031-9155/58/17/5983.

\bibitem{Giersch_Durst_2008}
J.~Giersch and J.~Durst,
\newblock Monte Carlo simulations in X-ray imaging,
\newblock Nuclear Instruments and Methods in Physics Research Section A:
  Accelerators, Spectrometers, Detectors and Associated Equipment {\bf 591},
  300–305 (2008),
\newblock doi:10.1016/j.nima.2008.03.078.

\bibitem{Unberath_Zaech_Lee_Bier_Fotouhi_Armand_Navab_2018}
M.~Unberath, J.-N. Zaech, S.~C. Lee, et al.,
\newblock DeepDRR -- A Catalyst for Machine Learning in Fluoroscopy-guided
  Procedures,
\newblock (2018),
\newblock doi:10.48550/arXiv.1803.08606.

\bibitem{Kusk_Jensen_Gram_Nielsen_Precht_2021}
M.~W.~Kusk, J.~M.~Jensen, E.~H.~Gram, J.~Nielsen, and H.~Precht,
\newblock Anode heel effect: Does it impact image quality in digital radiography? A systematic literature review,
\newblock Radiography {\bf 27}, 976–981 (2021),
\newblock doi:10.1016/j.radi.2021.02.014.

\bibitem{Abdurahman_Frysch_Bismark_Melnik_Beuing_Rose_2018}
S.~Abdurahman, R.~Frysch, R.~Bismark, S.~Melnik, O.~Beuing, and G.~Rose,
\newblock Beam Hardening Correction Using Cone Beam Consistency Conditions,
\newblock IEEE Transactions on Medical Imaging {\bf 37}, 2266–2277 (2018),
\newblock doi:10.1109/TMI.2018.2840343.

\bibitem{JIN2008124}
J.-Y. Jin, F.-F. Yin, S.~E. Tenn, P.~M. Medin, and T.~D. Solberg,
\newblock Use of the BrainLAB ExacTrac X-Ray 6D System in Image-Guided
  Radiotherapy,
\newblock Medical Dosimetry {\bf 33}, 124-134 (2008),
\newblock doi:10.1016/j.meddos.2008.02.005.

\bibitem{wang2004image}
Z.~Wang, A.~C.~Bovik, H.~R.~Sheikh, and E.~P.~Simoncelli,
\newblock Image quality assessment: from error visibility to structural
  similarity,
\newblock IEEE Transactions on Image Processing {\bf 13}, 600–612 (2004),
\newblock doi:10.1109/TIP.2003.819861.

\bibitem{lewis1995fast}
J.~P. Lewis,
\newblock Fast normalized cross-correlation,
\newblock Industrial Light \& Magic {\bf 10}, 1-5 (1995).

\bibitem{willmott2005advantages}
C.~Willmott and K.~Matsuura,
\newblock Advantages of the mean absolute error (MAE) over the root mean square
  error (RMSE) in assessing average model performance,
\newblock Climate Research {\bf 30}, 79–82 (2005),
\newblock doi:10.3354/cr030079.

\bibitem{Unberath_Zaech_Gao_Bier_Goldmann_Lee_Fotouhi_Taylor_Armand_Navab_2019}
M.~Unberath, J.-N. Zaech, C.~Gao, et al.,
\newblock Enabling machine learning in X-ray-based procedures via realistic
  simulation of image formation,
\newblock International Journal of Computer Assisted Radiology and Surgery {\bf
  14}, 1517–1528 (2019),
\newblock doi:10.1007/s11548-019-02011-2.

\bibitem{Ahmed_Garzanich_Melaragno_Nyirjesy_Windheim_Marquardt_Luttrull_Quails_VanKoevering_2024}
M.~Ahmed, M.~Garzanich, L.~E. Melaragno, et al.,
\newblock Exploring CT pixel and voxel size effect on anatomic modeling in
  mandibular reconstruction,
\newblock 3D Print Med. {\bf 10}, 21 (2024),
\newblock doi:10.1186/s41205-024-00223-0.

\bibitem{Pointon_Wen_Tugwell-Allsup_Sujar_Letang_Vidal_2023}
J.~L. Pointon, T.~Wen, J.~Tugwell-Allsup, A.~Sújar, J.~M. Létang, and F.~P.
  Vidal,
\newblock Simulation of X-ray projections on GPU: Benchmarking gVirtualXray
  with clinically realistic phantoms,
\newblock Computer Methods and Programs in Biomedicine {\bf 234}, 107500
  (2023),
\newblock doi:10.1016/j.cmpb.2023.107500.

\bibitem{Huang_Kurz_Freislederer_Manapov_Corradini_Niyazi_Belka_Landry_Riboldi_2024}
L.~Huang, C.~Kurz, P.~Freislederer, et al.,
\newblock Simultaneous object detection and segmentation for patient‐specific
  markerless lung tumor tracking in simulated radiographs with deep learning,
\newblock Medical Physics {\bf 51}, 1957–1973 (2024),
\newblock doi:10.1002/mp.16705.

\bibitem{Koetzier_Wu_Mastrodicasa_Lutz_Chung_Koszek_Pratap_Chaudhari_Rajpurkar_Lungren_et_al._2024}
L.~R. Koetzier, J.~Wu, D.~Mastrodicasa, A.~Lutz, et al.,
\newblock Generating Synthetic Data for Medical Imaging,
\newblock Radiology {\bf 312}, e232471 (2024),
\newblock doi:10.1148/radiol.232471.

\end{thebibliography}
